\documentstyle[aps,epsfig,prl]{revtex}

\begin{document}
\author{D. L. Pursey$^{1}$, A. M. Shirokov$^{2,3}$ and T. A. Weber$^{1}$}
\address{$^{1}$Department of Physics and Astronomy, \\
Iowa State University, Ames, Iowa 50011, USA.\\
$^{2}$International Institute for Theoretical and Applied Physics,\\
Iowa State University, Ames, Iowa 50011, USA.\\
$^{3}$Skobeltsyn Institute of Nuclear Physics, Moscow State University,\\
Moscow 119899, Russia\thanks{%
Permanent address.}.}
\title{Can one learn anything from proton-proton bremsstrahlung? Comment on H.~W.~
Fearing preprint nucl-th/9710061.}
\date{\today}
\maketitle

\begin{abstract}
Recently, Fearing (nucl-th/9710061) has argued that, as a matter of
principle, proton-proton bremsstrahlung can yield no more information about
the off-shell properties of the nucleon-nucleon interaction than can already
be deduced from the on-shell properties. In this note we challenge Fearing's
conclusion.
\end{abstract}

\bigskip

In his paper \cite{Fearing}, Fearing gives plausible but inconclusive
arguments that proton-proton bremsstrahlung can yield no more information
about the off-shell properties of the nucleon-nucleon interaction than can
already be deduced from the on-shell properties, and supports his
conclusions with a detailed calculation for $\pi ^{+}+\pi ^{0}\rightarrow
\pi ^{+}+\pi ^{0}+\gamma $, using a simple model Lagrangian involving only
the pion and photon fields. We have no quarrel with his analysis of this
latter process, but question the validity of drawing general conclusions
from that very simple special case. We also question the generality of
Fearing's discussion of off-shell effects in the nucleon-nucleon case.
Fearing's arguments are of two kinds. Firstly, he gives a general argument
that contributions to proton-proton bremsstrahlung arising from the
nucleon-nucleon interaction are inextricably mixed with contributions from
``contact terms'' which are in principle uncomputable within a potential
model of the NN interaction, thus leading to an unresolvable ambiguity. From
this he concludes that calculations of proton-proton bremsstrahlung using
only a potential model are incomplete and intrinsically unreliable.
Secondly, some off-shell contributions to the bremsstrahlung amplitude are
necessary consequences of the on-shell NN interaction. Fearing argues that
field transformations which leave the $S$-matrix unaltered affect any
additonal off-shell contributions, different from those which are necessary
consequences of the on-shell NN interaction, in an arbitrary manner,
allowing them to be eliminated entirely. Consequently he argues that
proton-proton bremsstrahlung can yield no information about off-shell
properties of the nucleon-nucleon interaction beyond those already implied
by the on-shell properties. We respond to these arguments in turn.

Fearing envisages computing the NN scattering amplitude both on-shell and
off-shell from some model (such as a potential model) of the NN interaction,
and using it (represented by shaded areas) in bremsstrahlung graphs such as
Fig. 1 (a)--(c). However there must also be contact terms represented by
Fig. 1(d) which cannot be correctly computed using only a potential model
for the NN interaction. Figure 2(a) shows an example of an irreducible
contact term. Furthermore, off-shell contributions to the nucleon propagator
in internal lines in diagrams Fig. 1(a)--(c) effectively collapse the line,
leading to a contribution similar to a contact term. Fearing therefore
asserts that off-shell contributions to the amplitude are inextricably
intermixed with irreducible contact terms which are intrinsically impossible
to compute in a potential model of the NN interaction. In response, we argue
that there is a significant difference between a (computable) contact term
arising from off-shell contributions to the internal proton propagators in
Figs. 1(a)--(c), modified so that the NN$\gamma $ vertex is fully
renormalized, and irreducible contact terms such as those illustrated by
Fig. 2(a). In a meson exchange theory of the nonrelativistic nucleon-nucleon
interaction, it is well known \cite{Gross} that crossed diagrams such as
Fig. 2(a) are less important than box diagrams, such as Fig. 2 (b), of the
same order. The contributions to the bremmstrahlung amplitude differ from
the contributions to elastic scattering from the corresponding diagrams only
by an additional propagator and an additional vertex. Therefore we expect to
find a similar result for bremsstrahlung provided at least the final state
protons are nonrelativistic. The singularity structure associated with
diagrams Fig. 2(a) and Fig. 2(b), examined using approximations similar to
those described by Gross \cite{Gross}, confirms this expectation. A
distorted wave approach is appropriate for low energy nucleons. Therefore
for the kinematical region in which the protons are nonrelativistic we
expect that irreducible point cointributions to the $pp\rightarrow pp\gamma $
amplitude will be suppressed and that a distorted wave Born approximation
(DWBA) should provide the dominant contribution. Recent work \cite
{moreppbremss} shows that such a kinematical region is of real interest.

More important are Fearing's arguments using field transformations. We are
interested with off-shell properties of the NN interaction which are {\em not%
} automatic consequences of the on-shell interaction. These must be ``EOM''
terms, i.e. terms which vanish in consequence of the equations of motion of
an assumed underlying field theory. One could in principle enumerate all
possible EOM contributions to the effective Lagrangian of the underlying
field theory, consistent with all necessary symmetries, and include them
with arbitrary coefficients analogous to the $\beta $'s of Fearing's pion
model. Now it is precisely EOM terms which are modified as a result of field
transformations \cite{Weinberg}. We next consider the most general field
transformation consistent with all of the symmetries of the theory. This
will also depend on a number of parameters, analogous to the $\alpha $'s of
Fearing's pion model. These field transformations will merely replace the
EOM terms in the Lagrangian by {\em the same} terms (since the most general
terms were already included) but with modified coefficients $\beta ^{\prime
}=\beta ^{\prime }\left( \alpha ,\beta \right) $. In the simple $\pi ^{+}\pi
^{0}$ bremsstrahlung model treated in detail by Fearing, given any $\beta
_{1}$ and $\beta _{2}$ it was possible to find $\alpha _{1}$ and $\alpha _{2}
$ such that $\beta _{1,2}^{\prime }=0$. However the most that can be
concluded from the general theory is that EOM terms in the Lagrangian
representing differing off-shell properties of the NN interaction can affect
observable processes only through a set of functions $\gamma _{i}\left(
\beta \right) $ which are invariant under the group of field
transformations. It is necessary to show that the set $\left\{ \gamma
\right\} $ of such functions is empty, and Fearing has not shown this for
the NN bremsstrahlung case.

Can we understand why the set $\left\{ \gamma \right\} $ need not be empty
for proton-proton bremsstrahlung even though $\left\{ \gamma \right\}
=\emptyset $ for the simple pion model treated by Fearing? We suggest that
Fearing's result is a consequence of the ``Cancellation theorem,'' stated on
p. 384 of Ref. \cite{Gross}. However this theorem does not hold for nucleons
interacting through the exchange of pions, because both nucleons and pions
carry isospin (see pp. 387--388 of Ref. \cite{Gross}).

Even if the set $\left\{ \gamma \right\} $ could be shown to be empty for a
particular field theory underlying the NN interaction, Fearing's argument
fails if the assumed underlying field theory is changed. The effective field
theory underlying the OBE potentials, based on nucleons and mesons, cannot
be changed by field transformations into a theory incorporating quarks in
the ``broken chiral symmetry'' phase, gluons, and Goldstone bosons,. While
there does not yet exist any NN potential rigorously derived from a
quark-gluon-Goldstone boson field theory, this approach inspired the
development of the ``Moscow potential'' \cite{Moscow}, which has been
successfully adapted to fit nucleon-nucleon properties. This model assumes
that the short range NN interaction is dominated by a 6-quark configuration
with orbital structure $s^{4}p^{2}$, which implies that the $s$-wave NN wave
function has a node at short distances. Recent work by Stancu {\it et al}. 
\cite{Glozman}, using a quark-Goldstone boson model which has shown success
in describing the hadron spectrum \cite{hadrons}, also requires the dominant
6-quark orbital structure to be $s^{4}p^{2}$, although the permutation
symmetry properties are different from those of Ref. \cite{Moscow}. One
should expect observable differences, due to differing off-shell properties
of the NN interaction, in the predictions for proton-proton bremsstrahlung
in treatments based on a nucleon-meson field theory or a
quark-gluon-Goldstone boson field theory, whether the calculation is
performed using field theoretical methods or potential models. Indeed,
recent distorted wave calculations \cite{moreppbremss} of the $pp\rightarrow
pp\gamma $ cross section, comparing the predictions of the Paris,
Hamada-Johnston, and Moscow potentials, show dramatic differences in the
cross section for incident (laboratory frame) energies of 350 MeV and 450
MeV in kinematic regions which correspond to the emission of hard photons.
Although an ambiguity in Ref. \cite{moreppbremss} arises from the difference
in the center of mass reference frames of the initial state and final state
protons, this is not as great as the difference between the predictions of
the Moscow and the other two potentials, and will be resolved by work in
progress \cite{Shirokov}.

In summary, we believe that Fearing has failed to demonstrate his thesis
that proton-proton bremsstrahlung can yield no information concerning the
off-shell properties of the nucleon-nucleon interaction. We argue that the
ambiguity in the contribution of contact terms noted by Fearing can be
minimized in a potential model by using distorted waves for the nucleons and
by suitably choosing the kinematical region. We also claim that Fearing has
not fully proved his formal results based on field transformation in the
case of proton-proton bremsstrahlung. Even if correct, however, his argument
cannot apply to potential models based on effective field theories which use
interpolating fields with different symmetry properties, such as quarks
versus nucleons.

One of us (A. M. S.) is grateful for hospitality at the International
Institute for Theoretical and Applied Physics of Iowa State University. We
also wish to thank Dr. G. Pivovarov for conversations and insightful
comments.

\begin{figure}
\centerline{\psfig{file=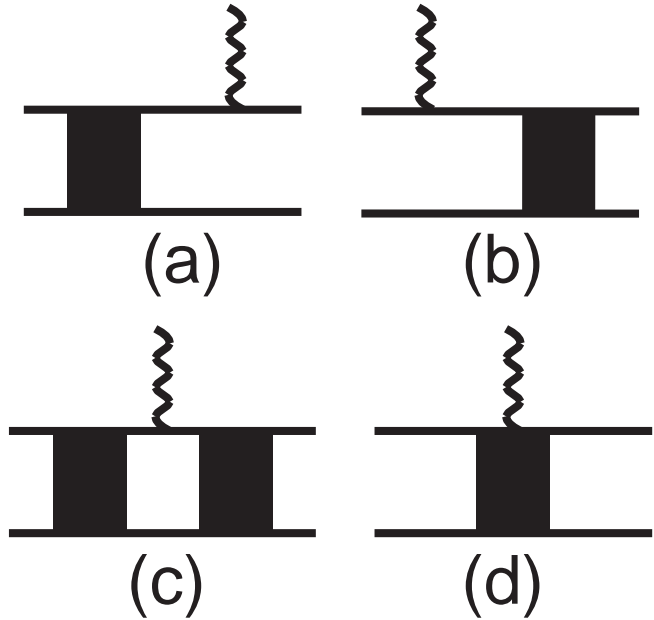, width=3.375in}}

\vspace{5mm}

\caption{Diagrammatic representation of contributions to proton-proton
bremsstrahlung. Off-shell effects contribute to the internal line
propagators in (a)--(c). Figure 1(d) represents contact terms.}
\end{figure}

\begin{figure}
\centerline{\psfig{file=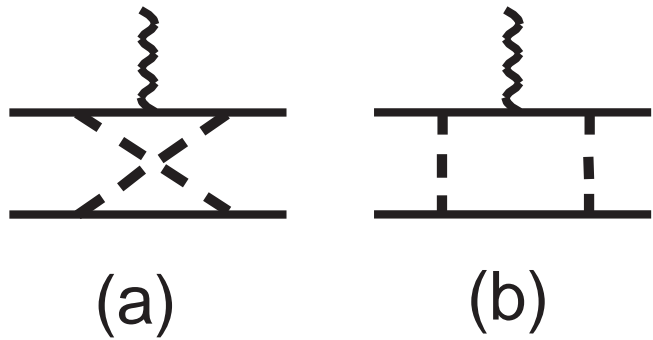, width=3.375in}}

\vspace{5mm}

\caption{(a) A typical irreducible ``contact term" diagram. (b) Box diagram for bremsstrahlung.}
\end{figure}


\begin{references}
\bibitem{Fearing}  Harold W. Fearing, ``Off-shell effects in nucleon-nucleon
bremsstrahlung,'' nucl-th/9710061, TRIUMF preprint TRI-PP-97-31 (1997).

\bibitem{Gross}  See Franz Gross, {\it Relativistic Quantum Mechanics and
Field Theory}, \S \S 12.1--12.2 (John Wiley \& Sons, Inc., 1993).

\bibitem{moreppbremss}  V. G. Neudatchin, N. A. Khokhlov, A. M. Shirokov,
and V. A. Knyr, Phys. Atomic Nuclei {\bf 60}, 971 (1997).

\bibitem{Weinberg}  S. Weinberg, {\it Quantum Theory of Fields}, vol. 1, p.
331 (Cambridge University Press, 1995).

\bibitem{Moscow}  V. G. Neudatchin, I. T. Obukhovsky, and Yu. F. Smirnov,
Phys. Lett. B {\bf 43}, 13 (1973); V. I. Kukulin and V. N. Pomerantsev,
Prog. Theor. Phys. {\bf 88}, 159 (1992), and papers cited therein.

\bibitem{Glozman}  Fl. Stancu, S. Pepin, and L. Ya. Glozman, Phys. Rev. C 
{\bf 56}, 2779 (1997).

\bibitem{hadrons}  L. Ya. Glozman, W. Plessas, K. Varga, and R. F.
Wagenbrunn, ``{\it Unified description of light- and strange-baryon spectra}%
,'' hep-ph/9706507.

\bibitem{Shirokov}  N. A. Khokhlov. Ph. D. thesis, Moscow State University
(1997); N. A. Khokhlov, V. G. Neudatchin, A. M. Shirokov, and V. A. Knyr, in
preparation.
\end{references}
\end{document}